\documentstyle[12pt,amsfonts]{article}
\parskip=\medskipamount

\newcommand {\cD}{{\cal D}}

\newcommand {\cF}{{\cal F}}

\newcommand {\cL}{{\cal L}}

\def\a{\alpha}
\def\b{\beta}

\def\d{\delta}
\def\e{\epsilon}

\def\k{\kappa}
\def\l{\lambda}

\def\q{\theta}

\def\s{\sigma}

\def\z{\zeta}

\def\F{\Phi}
\def\J{\Psi}

\def\O{\Omega}

\def\S{\Sigma}

\newcommand{\ad}{{\dot{\alpha}}}                           
\newcommand{\bd}{{\dot{\beta}}}                            
\newcommand{\ve}{\varepsilon}                            

\newcommand{\pa}{\partial}                           

\newcommand{\be}{\begin{equation}}
\newcommand{\ee}{\end{equation}}
\newcommand{\bea}{\begin{eqnarray}}
\newcommand{\eea}{\end{eqnarray}}
\newcommand{\non}{\nonumber}

\begin{document}
\begin{titlepage}

\begin{flushright}
ITP-UH-20/97\\
hep-th/9706169 \\
\end{flushright}

\begin{center}
\large{{\bf The Vector-Tensor Multiplet in Harmonic Superspace} } \\
\vspace{1.0cm}

\large{Norbert Dragon, Sergei M. Kuzenko\footnote{Alexander von
Humboldt
Research Fellow. On leave from Department of Quantum Field Theory,
Tomsk State University, Tomsk 634050, Russia.} and Ulrich Theis
} \\
\vspace{5mm}

\footnotesize{{\it Institut f\"ur Theoretische Physik, Universit\"at
Hannover\\
Appelstra{\ss}e 2, 30167 Hannover, Germany} \\
 }
\end{center}
\vspace{1.5cm}

\begin{abstract}
We describe the vector-tensor multiplet 
and derive its Chern-Simons coupling to the 
$N=2$ Yang-Mills gauge superfield in harmonic superspace.
\end{abstract}
\vspace{15mm}


\vfill
\null
\end{titlepage}
\newpage
\setcounter{footnote}{0}


\noindent
The $N=2$ vector-tensor multiplet, which was discovered many years ago by 
Sohnius, Stelle and West \cite{ssw} and then forgotten for a while, 
has recently received much interest \cite{wkll,cwfkst,cwft} due to
the fact that it originates in the low-energy effective Lagrangian of
$N=2$ heterotic string vacua. As a representation of $N=2$ supersymmetry, 
this multiplet is very similar to the massless $8+8$ Fayet-Sohnius 
hypermultiplet \cite{fss,s} which possesses an off-shell central charge
generating the equations of motion. 

The vector-tensor multiplet is the only known $N=2$, $D=4$ supersymmetric 
model that has not yet been formulated 
in the harmonic superspace \cite{gikos}.
Since the harmonic superspace is believed to be a universal framework
for $N=2$ supersymmetric theories, finding a relevant formulation 
for the vector-tensor multiplet seems to be of principal importance.
On the other hand, adequate formulations of the vector-tensor multiplet
in an $N=2$ superspace with central charges 
have been given in recent papers \cite{how,ghh}. Our primary goal in this 
letter is to show that  
the main results of Refs. \cite{how,ghh} have a natural origin in 
the harmonic superspace approach.
\vspace{5mm}

\noindent
We start with re-formulating the Sohnius prescription of constructing
supersymmetric actions \cite{s} in harmonic
superspace. The harmonic central charge superspace \cite{gikos}
extends the $N=2$ central charge superspace \cite{s}, 
with coordinates 
$\{ x^m,\, z,\, \theta_i^\alpha,\,\bar \theta^i_{\dot{\alpha}}\}$,  
$\overline{\theta_i^\alpha}=\bar \theta^{{\dot{\alpha}}\,i}$
(where $z$ is the central charge real variable), by the
two-sphere $S^2 =SU(2)/U(1)$ 
parametrized by harmonics, i.e. group
elements
\bea
&({u_i}^-\,,\,{u_i}^+) \in SU(2)\non\\
&u^+_i = \ve_{ij}u^{+j} \qquad \overline{u^{+i}} = u^-_i
\qquad u^{+i}u_i^- = 1 \;.
\eea
The analytic basis of the harmonic superspace defined by
\bea
& x^m_A = x^m - 2{\rm i} \q^{(i}\s^m {\bar \q}^{j)}u^+_i u^-_j \qquad
z_A = z + {\rm i}(\q^{+ \a} \q^-_\a - {\bar \q}^+_\ad {\bar \q}^{- \ad})
\non \\
& \q^\pm_\a=u^\pm_i \q^i_\a \qquad {\bar \q}^\pm_{\dot\a}=u^\pm_i{\bar
\q}^i_{\dot\a} 
\eea
is most suitable to the description of analytic superfields $\F(\z, u)$
which depends only on the variables
\be
\z^M \equiv\{x^m_A,z_A,\q^{+\a},{\bar\q}^+_{\dot\a}\} 
\ee
and harmonics $u^\pm_i$ (the original basis of the harmonic superspace
is called central \cite{gikos}). Below we will mainly work in the analytic
basis and omit the corresponding subscript ``$A$''. The explicit
expressions for the covariant derivatives 
$D^\pm_\alpha = D^i_\alpha u^\pm_i $,
${\bar D}^\pm_{\dot\alpha}={\bar D}^i_{\dot\alpha} u^\pm_i$
in the analytic basis can be found in Ref. \cite{gikos}.

The GIKOS rule \cite{gikos} of constructing $N=2$ supersymmetric
actions
\be
\int d\z^{(-4)} du \, \cL^{(4)} \qquad \qquad
d\z^{(-4)}=d^4x d^2\q^+d^2{\bar \q}^+
\label{1}
\ee
involves an analytic superfield $\cL^{(4)}(\z, u)$ of $U(1)$-charge $+4$
which is invariant (up to derivatives) under central charge transformations
generated by $\pa_z \equiv \pa / \pa z$
\be 
\frac{\pa}{\pa z} \cL^{(4)} = \frac{\pa}{\pa x^m} f^{(4)m}\;.
\label{2}
\ee
Here $\cL^{(4)}$ is a function of the dynamical superfields, their 
covariant derivatives and, in general, of the harmonic variables. 

In harmonic superspace there exists a prescription to construct
invariant actions even for non-vanishing 
central charges. The construction makes  use of a constrained
analytic superfied $\cL^{++}(\z, u)$.
$\cL^{++}$ is an analytic superfield of $U(1)$-charge $+2$
\be
D^+_\a \cL^{++} = {\bar D}^+_\ad \cL^{++} = 0
\label{anal}
\ee
which satisfies the covariant constraint
\be 
D^{++}_{\rm c} \cL^{++} = 0\;.
\label{3}
\ee
$D^{++}_{\rm c}$ acts according to
\be
D^{++}_{\rm c} = D^{++} + {\rm i}\, 
(\q^{+ \a} \q^+_\a - {\bar \q}^+_\ad {\bar \q}^{+ \ad}) 
\frac{\pa}{\pa z} \qquad
D^{++} = u^{+i} \frac{\pa}{\pa u^{-i}} - 2{\rm i}\,\q^+ \s^m {\bar \q}^+
\frac{\pa}{\pa x^m}
\label{4}
\ee
on analytic superfields. Then the action 
\be
S = \int d\z^{(-4)} du \, 
\left( (\q^+)^2  - ({\bar \q}^+)^2 \right) \cL^{++}
\label{5}
\ee
is supersymmetric and, hence, invariant under
central charge transformations. Under a supersymmetry transformation
\bea
\d x^m &=& -2{\rm i}\, (\e^- \s^m {\bar \q}^+ + \q^+ \s^m {\bar \e}^-)
\non \\
\d z &=& 2{\rm i}\, (\e^- {\bar \q}^+ - {\bar \e}^- {\bar \q}^+)
\non \\
\d \q^+_\a &=& \e^+_\a \qquad \d {\bar \q}^+_\ad = {\bar \e}^+_\ad 
\label{6}
\eea
$S$ changes by
\be
\d S = \int d\z^{(-4)} du \, \left\{ 
\left((\q^+)^2 - ({\bar \q}^+)^2 \right)
\d z \frac{\pa}{\pa z} \cL^{++}
- 2 \left( \e^+ \q^+ - {\bar \e}^+ {\bar \q}^+ \right) \cL^{++}
\right\} \;.
\label{7}
\ee
Making use of the identity 
$2 \left( \e^+ \q^+ - {\bar \e}^+ {\bar \q}^+ \right)
=- {\rm i}\, D^{++}  \d z $
and integrating by parts in (\ref{7}), one arrives at
\be
\d S = -{\rm i}\,\int d\z^{(-4)} du \, \d z D^{++}_{\rm c} \cL^{++}
\label{8}
\ee
and this is equal to zero due to (\ref{3}). The action (\ref{5})
is real if $\cL^{++}$ is imaginary
\be
\breve{\cL}^{++} = - \cL^{++}
\label{9}
\ee
with respect to the analyticity preserving conjugation (smile)
$\; \breve{} \; = \;\stackrel{\star}{\bar{}}$ introduced in \cite{gikos}, 
where the operation $\,\bar{}\,$ denotes the complex conjugation 
and the operation ${}^\star$ is defined by $(u^+_i)^\star = u^-_i$,
$(u^-_i)^\star = - u^+_i$, hence $(u^{\pm}_i)^{\star \star} = - u^{\pm}_i$.

Eq. (\ref{5}) is the formulation of
the Sohnius action  \cite{s} (see also \cite{hst}) in the harmonic superspace. 
More explicitly, in the central basis the
constraint (\ref{3}) means
\be
\cL^{++} = \cL^{ij} (x,z,\q) u^+_i u^+_j
\ee
for some $u$-independent superfields $\cL^{ij}$, and the analyticity
conditions (\ref{anal}) take the form
\be
D^{(i}_\a \cL^{jk)} = {\bar D}^{(i}_\ad \cL^{jk)} = 0\;.
\ee
Since
\be
 d\z^{(-4)}  =  \frac{1}{16} d^4 x D^{-\a}D^-_\a 
{\bar D}^-_\ad {\bar D}^{-\ad}    
\ee
the action (\ref{5}) turns, upon integrating over $S^2$, into
\be
S = \frac{1}{12}\, \int d^4 x \,\left( D^{\a i} D_\a^j
- {\bar D}_\ad^i {\bar D}^{\ad j} \right) \cL_{ij}\;.
\ee

It is instructive to consider two examples.
The $8+8$ Fayet-Sohnius off-shell hypermultiplet coupled to the $N=2$
gauge multiplet is described in the $N=2$
central charge superspace by a superfield $q_i (x,z,\q)$ satisfying
the constraints \cite{s}
\be
\cD^{(i}_\a q^{j)} = {\bar \cD}^{(i}_\ad q^{j)} = 0
\ee
with $\cD_\a^i$ the gauge covariant derivatives.
This is equivalent to the fact that 
the superfield $q^+= q_i u^{+i}$ is covariantly analytic
\be
\cD^+_\a q^+ = {\bar \cD}^+_\ad q^+ = 0
\ee
and satisfies the gauge covariant constarint
\be
\cD^{++}_{\rm c} q^+ = 0\;.
\label{rr}
\ee
In the analytic basis $\cD^{++}_{\rm c}$ is
\be
\cD^{++}_{\rm c} = \cD^{++} + {\rm i}\,
(\q^{+ \a} \q^+_\a - {\bar \q}^+_\ad {\bar \q}^{+ \ad})
\frac{\pa}{\pa z} \qquad \cD^{++} = D^{++} + {\rm i}\,V^{++}
\ee
where $V^{++}$ is the analytic Yang-Mills gauge prepotential \cite{gikos}.
Therefore, the gauge invariant superfield 
\be
\cL^{++}_{{\rm FS}} = \frac{{\rm i}}{2} \left(\breve{q}^+ \pa_z q^+
- \pa_z \breve{q}^+ q^+\right) +m \,\breve{q}^+ q^+ 
\ee
meets the requirements (\ref{anal}) and (\ref{3}).
Because of (\ref{rr}), the corresponding action
can be rewritten in the following form
\be
S_{{\rm FS}} = \int d\z^{(-4)} du \,
\left\{ -\breve{q}^+ \cD^{++}q^+ + m \,\breve{q}^+ q^+ \left(
(\q^+)^2  - ({\bar \q}^+)^2 \right) \right\}
\ee
which is very similar to the action functional of the infinite-component
$q$-hypermultiplet \cite{gikos}. 
Another non-trivial example is the effective action of the $N=2$
super Yang-Mills theory \cite{gates,seiberg}
(supersymmetry without central charges)
\be
S_{{\rm SYM}} = {\rm tr}\,\int d^4 x d^4\q \cF(W) +
{\rm tr}\,\int d^4 x d^4{\bar \q} {\bar \cF}({\bar W})
\ee
where $W$ is the covariantly chiral field strength of the $N=2$ gauge
superfield \cite{gsw}. $S_{{\rm SYM}}$ can be represented as follows
\bea
S_{{\rm SYM}} &=& \frac{1}{4}\,{\rm tr}\,\int d\z^{(-4)} du \,
\left( (\q^+)^2  - ({\bar \q}^+)^2 \right) \cL^{++}_{{\rm SYM}} \non\\
\cL^{++}_{{\rm SYM}} &=& \left(\cD^+\right)^2 \cF(W) -
\left({\bar \cD}^+\right)^2 {\bar \cF}({\bar W}) \;.
\eea
It is obvious that $\cL^{++}_{{\rm SYM}}$ satisfies the requirements
(\ref{anal}) and (\ref{3}).  
\vspace{5mm}

\noindent
A free vector-tensor multiplet can be described in the harmonic superspace
by an analytic spinor superfield $\J^+_\a (\z ,u)$
\be
D^+_\a \J^+_\b = {\bar D}^+_\ad \J^+_\b = 0
\label{3.1}
\ee
subject to the constraints
\bea
& D^{++}_{{\rm c}} \J^+_\a = 0
\label{3.2} \\
& D^{-\a} \J^+_\a = {\bar D}^{-\ad} \breve{\J}^+_\ad
\label{3.3}
\eea
with $\breve{\J}^+_\ad$ the smile-conjugate of $\J^+_\a$.
Eq. (\ref{3.2}) implies that in the central basis $\J^+_\a$ reads
\be
\J^+_\a  = \J_{\a i} (x,z,\q) u^{+i} \qquad
\breve{\J}^+_\ad  = -{\bar \J}^i_\ad (x,z,\q) u^+_i
\label{3.4}
\ee
for some $u$-independent superfields $\J_{\a i}$ and its complex
conjugate ${\bar \J}^i_\ad$ . Then, the analyticity
conditions (\ref{3.1}) are equivalent to
\be 
D^{(i}_\a \J^{j)}_\b = {\bar D}^{(i}_\ad \J^{j)}_\b = 0 
\label{3.5}
\ee
and the reality condition (\ref{3.3}) takes the form 
\be
D^{\a i} \J_{\a i} = {\bar D}_{\ad i} {\bar \J}^{\ad i} \;.
\label{3.6}
\ee
Eqs. (\ref{3.5}) and (\ref{3.6}) constitute the constraints
defining the field strengths of the free vector-tensor multiplet \cite{how}.

Using the anticommutation relations 
\bea
& \{ D^+_\a, D^-_\b \} =2{\rm i}\, \ve_{\a \b} \pa_z \qquad
\{ {\bar D}^+_\ad, {\bar D}^-_\bd \} =2{\rm i}\, \ve_{\ad \bd} \pa_z
\non \\
& \{ D^+_\a, {\bar D}^-_\bd \} = - \{ D^-_\a, {\bar D}^+_\bd \}
= - 2{\rm i}\,\pa_{\a \bd}
\label{3.7}
\eea
one immediately deduces from (\ref{3.1}) and (\ref{3.3})  generalized
Dirac equations
\be
\pa_z \J^+_\a = - \pa_{\a \bd} \breve{\J}^{+\bd} \qquad
\pa_z \breve{\J}^+_\ad =  \pa_{\b \ad} \J^{+\b}
\label{3.8}
\ee
and hence
\be
\pa_z^2 \J^+_\a = \Box \J^+_\a \;.
\label{3.9}
\ee
The last relation can be also obtained from (\ref{3.1}) and 
(\ref{3.2}), in complete analogy to the Fayet-Sohnius hypermultiplet.
We read eqs. (\ref{3.8}) and (\ref{3.9}) as a definition of the
central charge. If one had not allowed for a central charge then
the constraints (\ref{3.1})--(\ref{3.3}) would have restricted
the multiplet to be on-shell.

The super Lagrangian associated with the vector-tensor multiplet reads
\be 
\cL^{++}_{\rm vt,free} = - \frac{1}{4} \left( \J^{+ \a} \J^+_\a -
\breve{\J}^+_\ad \breve{\J}^{+\ad} \right)\;.
\label{3.10}
\ee
Under central charge transformations it changes by derivatives
\be
\pa_z \cL^{++}_{\rm vt,free} = - \frac{1}{2} \pa_{\a \ad} 
\left(\J^{+\a} \breve{\J}^{+\ad} \right) \;.
\label{3.11}
\ee
The functional
\be
S_{{\rm vt,free}} = \int d\z^{(-4)} du \, 
\left( (\q^+)^2  - ({\bar \q}^+)^2 \right) \cL^{++}_{{\rm vt,free}}
\label{3.12}
\ee
can be seen to coincide with the action given in \cite{how}.
Another possible structure
\be
\cL^{++}_{\rm der,free} = - \frac{\rm i}{4} \left(\J^{+ \a} \J^+_\a +
\breve{\J}^+_\ad \breve{\J}^{+\ad} \right)
\label{3.13}
\ee
produces a total derivative when integrated over the superspace.
\vspace{5mm}

\noindent
The constraints (\ref{3.1})--(\ref{3.3}) can be partially solved
in terms of a real $u$-independent potential $L(x,z,\q)$ 
\be
\J^+_\a = {\rm i}\, D^+_\a L \qquad  \qquad {\bar L} = L  
\label{4.1}
\ee
which is still restricted by
\be
D^+_\a D^+_\b L = D^+_\a {\bar D}^+_\bd L= 0 \;.
\label{4.2}
\ee

If one includes coupling to the $N=2$ Yang-Mills gauge superfield, described
by the covariantly chiral strength $W$ and its conjugate $\bar W$
\cite{gsw}, the constraints (\ref{4.2}) can be consistently deformed
as follows \cite{ghh}
\bea
\cD^{+ \a} \cD^+_\a {\Bbb L} &=& \k\, {\rm tr}\, \left( {\bar \cD}^+_\ad 
{\bar W}
\cdot {\bar \cD}^{+ \ad} {\bar W} \right) \label{4.3} \\
\cD^+_\a {\bar \cD}^+_\bd {\Bbb L} &=& - \k\, {\rm tr}\, \left( \cD^+_\a W \cdot
{\bar \cD}^+_\bd {\bar W} \right)
\label{4.4}
\eea 
where $\Bbb L$ is a real $u$-independent gauge invariant superfield while
$W$ is invariant under the central charge.
Such a deformation corresponds in particular to the Chern-Simons coupling 
of the antisymmetric tensor field, contained in the vector-tensor multiplet,
to the Yang-Mills gauge field. 

The independent components of the vector-tensor multiplet can be chosen as
\bea
& \F = {\Bbb L}\,|  \qquad D = \pa_z {\Bbb L}\,| \non \\
& \psi^i_\a = \cD^i_\a {\Bbb L}\,| \qquad
\bar{\psi}_{\ad i} = \bar{\cD}_{\ad i}
{\Bbb L}\,| \non \\
& G_{\a\b} = \frac{1}{2} [ \cD_{\a i}\, , \cD_\b^i ]\,
{\Bbb L}\,| \qquad \bar{G}_{\ad\bd} = -\frac{1}{2} [ \bar{\cD}_{\ad i}\,
, \bar{\cD}_{\bd}^i ]\, {\Bbb L}\,| \non \\
& H_{\a\ad} = \bar{H}_{\a\ad}  =  - \frac{1}{2} [ \cD^i_\a\, ,\,
\bar{\cD}_{\ad i} ]\, {\Bbb L}\,|
\eea
while the components of the vector multiplet are
\bea
& X  =  W\,| \qquad \bar{X} = \bar{W}\,| \non \\ 
& \l^i_\a = \cD^i_\a W\,| \qquad \bar{\l}_{\ad i} = \bar{\cD}_{\ad i}
\bar{W}\,|\non \\
& \quad F_{\a\b} = -\frac{1}{4} [
\cD_{\a i}\, , \cD_\b^i ] W\,| \qquad \bar{F}_{\ad\bd} = \frac{1}{4} [
\bar{\cD}_{\ad i}\, , \bar{\cD}_{\bd}^i ] \bar{W}\,| \non \\
& Y^{ij} = -\frac{\rm i}{4} \big( \cD^{\a i} \cD_\a^j W +
\bar{\cD}_\ad^i \bar{\cD}^{\ad j} \bar{W} \big)\,|
\eea
with $F_{mn}$ the field strength associated with the Yang-Mills gauge field
$A_m$. The fields $H_m$ and $G_{mn}$ are subject to the constraints
\bea
\pa_m H^m & = & \k\, {\rm tr}\, \big\{ F^{mn} \tilde{F}_{mn} - \frac{1}{2}
\pa_m (\l^i \s^m \bar{\l}_i) \big\} \non \\
\pa_m \tilde{G}^{mn} & = & 2 \k\, \pa_m\, {\rm tr}\, \big\{ (X+\bar{X})
\tilde{F}^{mn} + \frac{\rm i}{4} \l^i \s^{mn} \l_i + \frac{\rm i}{4}
\bar{\l}_i \bar{\s}^{mn} \bar{\l}^i \big\}
\eea
which can be solved in terms of an antisymmetric tensor $B_{mn}$ and
a vector $V_m$
\bea
H^m & = & \ve^{mnkl} \pa_n B_{kl} + \k\, {\rm tr}\, \big\{ \ve^{mnkl} (A_n
F_{kl} - \frac{2}{3} A_n A_k A_l) - \frac{1}{2} \l^i \s^m \bar{\l}_i
\big\}\non \\
G_{mn} & = & \pa_m V_n - \pa_n V_m + 2\k\, {\rm tr}\, \big\{
(X+\bar{X}) F_{mn} + \frac{1}{4} \l^i \s_{mn} \l_i - \frac{1}{4} \bar{\l}_i
\bar{\s}_{mn} \bar{\l}^i \big\}\; .
\eea

Because of the constraints (\ref{4.3}) and (\ref{4.4}), the superfield
$\cD^+_\a {\Bbb L}$ is no longer analytic.
But also the Lagrangians
(\ref{3.10}) and (\ref{3.13}) can be deformed to obtain supersymmetric
actions with Chern-Simons interactions.
Similarly to Ref. 
\cite{ghh}, let us introduce the following real superfield
\be 
\S = {\Bbb L} -\frac{\k}{2} \,{\rm tr}\,\left(W - {\bar W}\right)^2 \;.
\label{4.5}
\ee   
Using the Bianchi identities \cite{gsw,s2}
\be
{\bar \cD}^+_\ad W = 0 \qquad
(\cD^+)^2 W = ({\bar \cD}^+)^2 {\bar W}
\label{4.6}
\ee
one can prove the important identities
\bea
\cD^+_\a {\bar \cD}^+_\bd \S &=& 0 \label{4.7} \\
(\cD^+)^2 \S &=& - ({\bar \cD}^+)^2 \S \non \\
&= &\frac{\k}{2} \left\{ ({\bar \cD}^+)^2 {\rm tr}\,\left({\bar W}^2 \right)
- (\cD^+)^2 {\rm tr}\,\left( W^2 \right) \right\}\;.
\label{4.8}
\eea
Therefore, the imaginary superfield
\be
\cL^{++}_{\rm vt} = \frac{1}{4} \left\{ \cD^{+ \a} \S \cD^+_\a \S
+\S (\cD^+)^2 \S - {\bar \cD}^+_\ad \S {\bar \cD}^{+ \ad} \S \right\}
\label{4.9}
\ee
satisfies both the
constraints (\ref{anal}) and (\ref{3}), and therefore can be used 
to construct a supersymmetric action. 
The corresponding action functional obtained by the rule (\ref{5}) 
describes the Chern-Simons coupling of the vector-tensor multiplet to the
$N=2$ gauge multiplet. It was first derived in component
approach \cite{cwft} and then in $N=2$ superspace \cite{ghh}.
We give only the bosonic part of the component Lagrangian:
\bea
\cL_{\rm vt} & = & \frac{1}{2} \pa^m \F\, \pa_m \F - \frac{1}{2}
H^m H_m - \frac{1}{4} G^{mn} G_{mn} + \frac{1}{2} D^2 \non \\
& & + {\rm i}\k \,
G^{mn}\, {\rm tr}\, \big\{ (X-\bar{X}) \tilde{F}_{mn} \big\} 
 - {\rm i}\k\, H^m\, {\rm tr}\, \big\{ (X-\bar{X}) \cD_m (X+\bar{X})
\big\} \non \\
& &- 2\k \left( \F - \frac{\k}{2} {\rm tr}\, (X-\bar{X})^2 \right) 
 {\rm tr}\, \big\{ \cD^m \bar{X}\, \cD_m X - \frac{1}{2} F^{mn}
F_{mn} - \frac{1}{2} Y^{ij} Y_{ij} + \frac{1}{4} [ X ,\bar{X} ]^2
\big\} \non \\
&   & + 2 \k^2\,  {\rm tr}\, \big\{ (X-\bar{X}) \cD^m \bar{X}
\big\}\, {\rm tr}\, \big\{ (X-\bar{X}) \cD_m X \big\} \non \\
& &- \k^2\, {\rm tr}\,
\big\{ (X-\bar{X}) F^{mn} \big\} 
{\rm tr}\, \big\{ (X-\bar{X}) F_{mn} \big\} \non \\
& &- \k^2\,
{\rm tr}\, \big\{ (X-\bar{X}) Y^{ij} \big\}\, {\rm tr}\, 
\big\{ (X-\bar{X}) Y_{ij}
\big\}  
+  \mbox{fermionic terms}\; . 
\eea

Now, we generalize the total derivative Lagrangian (\ref{3.13})
(which is an $N=2$ analog of $\tilde{F}\,F$ or $\q$-term). 
Similar to \cite{ghh}, we introduce
the real superfield
\be
\O = {\Bbb L} + \frac{\k}{2} \,{\rm tr}\,\left(W + {\bar W}\right)^2 \;. 
\label{4.11}
\ee
Its properties read
\bea
\cD^+_\a {\bar \cD}^+_\bd \O &=& 0 \label{4.12} \\
(\cD^+)^2 \O &=&  ({\bar \cD}^+)^2 \O \non \\
&= &\frac{\k}{2} \left\{ ({\bar \cD}^+)^2 {\rm tr}\,\left({\bar W}^2 \right)
+ (\cD^+)^2 {\rm tr}\,\left( W^2 \right) \right\}\;.
\label{4.13}
\eea
As a consequence, the imaginary superfield
\be
\cL^{++}_{\rm der} = \frac{\rm i}{4} \left\{ \cD^{+ \a} \O \cD^+_\a \O
+\O (\cD^+)^2 \O + {\bar \cD}^+_\ad \O {\bar \cD}^{+ \ad} \O \right\}
\label{4.14}
\ee
respects both the constraints (\ref{anal}) and (\ref{3}) and therefore
defines a supersymmetric action.

The Lagrangian (\ref{4.14}) is the deformation of (\ref{3.13}).
It is therefore a deformation of the Chern-Simons form $\tilde{F}\,F$
which carries topological information. 
In components, the bosonic Lagrangian reads
\bea
\cL_{\rm der,bos} & = & \pa_m \Big[ \left( \F + \frac{\k}{2} {\rm tr}\,
(X+\bar{X})^2 \right) \left( H^m - {\rm i}\k\, {\rm tr}\, \big\{
(X+\bar{X}) \cD^m (X-\bar{X}) \big\} \right) \non \\
 &   & \qquad + \frac{1}{2} \ve^{mnkl} V_n \pa_k V_l\, \Big]\, .
\eea
and contains total derivative terms only.

In summary, in the present paper we have described the vector-tensor
multiplet and its Chern-Simons coupling to the $N=2$ gauge multiplet
in harmonic superspace. It would be of interest to find an unconstrained
prepotential superfield formulation for the vector-tensor multiplet,
which may exist, similar to the Fayet-Sohnius hypermultiplet,
in harmonic superspace only.

\vspace{5mm}

\noindent
After this work had appeared on the hep-th archive we became aware of 
a recent paper \cite{bho} where the authors presented a two-form formulation
of the vector-tensor multiplet in central charge superspace and derived
its coupling to the non-Abelian supergauge multiplet via the Chern-Simons
form. The later paper is a natural development of the research started 
in \cite{how}.

\vspace{1cm}

\noindent
{\bf Acknowledgements.}
This work was supported by
the RFBR-DFG project No 96-02-00180,
the RFBR project No 96-02-16017 and by the Alexander
von Humboldt Foundation.


\end{document}